\documentclass[amsmath,amssymb,aps,prd,twocolumn,superscriptaddress]{revtex4}

\usepackage{amsmath,amssymb,amsfonts}
\usepackage[dvips]{graphicx}
\usepackage{color}

\newcommand{\comment}[1]{}

\newcommand{\Hess}{{\cal H}}
\newcommand{\Edns}{{\varepsilon}}
\newcommand{\VEV}[1]{\left\langle{#1}\right\rangle}
\newcommand{\PSfig}[2]{\includegraphics[width=#1]{Figs/#2}}
\newcommand{\nn}{\nonumber\\}

\newcommand{\LLE}{\lambda^\mathrm{LLE}}

\begin{document}

\title{Chaotic behavior in classical Yang-Mills dynamics}

\author{Teiji Kunihiro}
\affiliation{Department of Physics, Kyoto University, Kyoto 606-8502, Japan}
\author{Berndt M\"uller}
\affiliation{Department of Physics \& CTMS, Duke University, Durham, NC 27708, USA}
\author{Akira Ohnishi}
\affiliation{Yukawa Institute for Theoretical Physics, Kyoto University,
Kyoto 606-8502, Japan}
\author{Andreas Sch\"afer}
\affiliation{Institut f\"ur Theoretische Physik, Universit\"at Regensburg,
D-93040 Regensburg, Germany}
\affiliation{Yukawa Institute for Theoretical Physics, Kyoto University,
Kyoto 606-8502, Japan}
\author{Toru T. Takahashi}
\affiliation{Yukawa Institute for Theoretical Physics, Kyoto University,
Kyoto 606-8502, Japan}
\author{Arata Yamamoto}
\affiliation{Department of Physics, Kyoto University, Kyoto 606-8502, Japan}

\date{\today}
\begin{abstract}
Understanding the underlying mechanisms causing rapid thermalization deduced
for high-energy heavy ion 
collissons is still a challenge. To estimate the thermalization time,
entropy growth for classical Yang-Mills theories is studied, based on 
the determination of Lyapunov exponents. Distinct regimes for short,
medium  and long sampling times are characterized by different properties 
of their spectrum of Lyapunov exponents. 
Clarifying the existence of these regimes and their implications 
for gauge-field dynamics is one of the results of this contribution. 
As a phenomenological application we conclude that for pure gauge 
theories with random initial conditions thermalization occures 
within few fm/c, an estimate which can 
be reduced by the inclusion of fermions, specific initial conditions etc. 
\end{abstract}
\maketitle

\section{Introduction}
\label{Sec:Introduction}

Experiments have shown that a new form of strongly interacting matter
with very high energy density and unusual transport properties is created in
collisions between heavy nuclei at energies attainable at the Relativistic
Heavy Ion Collider (RHIC), up to 200 GeV per nucleon pair in the center of
mass \cite{Whitepapers}. 
Theoretical arguments as well as circumstantial experimental evidence
suggest that this matter is a strongly coupled quark-gluon plasma
\cite{Gyulassy:2004zy}.
The early thermalization of this matter leading to the formation of a quark-gluon
plasma is one of the largest unexplained puzzles in RHIC physics.
Hydrodynamic simulations are consistent with a thermalization time
of 1.5 fm/c or less \cite{heinz}.
It is generally believed that the instability and consequent exponential growth
of intense gluon fields would be the origin of early thermalization.
Various plasma instabilities such as the Weibel instability~\cite{Mrowczynski:1993qm}
and the Nielsen-Olesen instability~\cite{Nielsen:1978rm} 
can cause the exponential growth of the amplitude of unstable
modes of the SU(3) gauge field.
The plasma instability may be characterized by the negative curvature
of the potential, leading to the equation of motion
\begin{equation}
\ddot{X}_i=\lambda_i^2 X_i\ ,
\end{equation}
where $X_i$ denotes the field variable in the unstable mode.
The energy stored in the intense gauge field eventually produces 
abundant particles and evolves towards a thermalized state.
The thermalization mechanism governing this transition is 
not yet clear, and the time scale on which it occurs is not known. 
The equilibration problem is simplified, however, by the high
occupation probability of the unstable modes, which makes a
quasi-classical treatment of the thermalization process, at least
of its initial stages, possible.

In the classical dynamics, the apparent entropy of an isolated 
system is produced by the increasing complexity in phase space.
The distance between classical trajectories starting from very similar initial 
conditions grows exponentially in the long-time evolution of a chaotic system,
\begin{equation}
\label{eq:LE}
|\delta X_i(t)| \propto e^{\lambda_i t}\ ,
\end{equation}
where $\delta X_i$ represents the separation of trajectories,
and $\lambda_i$ is referred to as the Lyapunov exponent (LE).
The entropy production rate is given by the Kolmogorov-Sina\"i (KS) entropy,
which is defined as the sum of positive LEs,
$dS/dt=S_{KS} \equiv \sum_{\lambda_i>0} \lambda_i$.
The production of entropy at the quantum level poses additional problems
such as the decoherence of the quantum state of the system~\cite{Fries:2008vp},
since the evolution in pure state generates no entropy 
and some kind of coarse graining is necessary.
Kunihiro, M\"uller, Ohnishi, and Sch\"afer \cite{Kunihiro:2008gv} 
(henceforth referred to as KMOS) proposed to apply the Husimi function,
a smeared Wigner function with minimal wave packets, to define a minimally
coarse grained entropy, the Wehrl entropy, and showed that it grows at the 
rate of the KS entropy in the classical long-time limit, i.e. if the system has enough 
time to sample the complete phase space.
In that paper application was limited to simple cases where the number of degrees 
of freedom is essentially one. 
In this contribution we extend the KMOS framework to
more realistic processes.

In this work, we analyze the chaotic behavior in the classical Yang-Mills
(CYM) evolution.
Specifically, we analyze the exponentially growing behavior
of the distances between the trajectories.
We find that we have to distinguish different regimes, depending on sampling time,
namely a kinetic stage for short sampling times, an intermediate- and 
a long-time regiem.In each case 
we consider the exponential growth rate 
of the distance between two trajectories, which follows the equation of motion,
\begin{align}
\delta \dot{X}(t)={\cal H}(t,X)\delta X(t)
\end{align}
where ${\cal H}$ is the so-called Hesse matrix or {\em Hessian}
and analyse the time evolution of the distance vector $\delta X$.\\ 
i) The instantaneous change of $\delta X$ is determined by the eigenvalues 
of the Hessian, which we will refer to as the {\em local Lyapunov exponents} (LLE).
\\
ii)
The evolution of the distance on ergodic time scales is described by the
Lyapunov exponents (\ref{eq:LE}), which we will refer to as {\em global
Lyapunov exponents} (GLE). 
\\
iii)
For the third, intermediate time period,
the Hessian changes due to the nonlinear coupling 
among the different field modes, but the energy remains localized among
the primary unstable modes. By using the Trotter formula, 
we can numerically integrate the equation of motion for the tangent space
$\delta X$, and construct the time-evolution matrix for an intermediate time period.
We will refer to the eigenvalues of the time-evolution matrix as {\em intermediate Lyapunov exponents} (ILE).

Because the ILEs describe the evolution of the strongly excited Yang-Mills 
field modes during the time when the field configuration is still far away from 
equilibrium and a quasi-classical description of the dynamics of the Yang-Mills
field is appropriate, the ILEs are the most relevant Lyapunov exponents for 
the early thermalization at RHIC.

Below we obtain the distribution of these three kinds of Lyapunov exponents.
Since they govern the growth rate of the coarse grained entropy residing in
the Yang-Mills field, they will allow us to estimate the equilibration time as 
$\tau_{eq}\simeq \Delta S/S_{KS}$,
where $\Delta S$ is the increase of entropy necessary for equilibration.

Since classical  CYM theory has no conformal anomaly (it does not know about
$\Lambda_{QCD}$)
all statistical quantities should scale like
$\varepsilon^{n/4}$, where $\varepsilon$ is the energy density
and $n$ is the mass dimension of that quantity.
For example, the KS entropy has the mass dimension and scales
as $S_\mathrm{KS} \propto \varepsilon^{1/4}$.

For the initial stage of high energy heavy ion collisions 
the relevant scale is the saturation scale $Q_s$, which is related to
the initial energy density in the color glass condensate (CGC) model
as $\varepsilon = Q_s^4/g^2$, implying that the time scale of very early dynamics 
is given by $1/Q_s$.\\

Not surprisingly, 
Fries, M\"uller and Sch\"afer have indeed found that decoherence (which is 
probably the fastest mechanism for entropy production)
happens indeed on this time scale~\cite{Fries:2008vp}.

However, they also found that decoherence can only generate a fraction of the entropy needed
to justify a hydrodynamic treatment.

The real-time gauge field dynamics discussed in this contribution is treated 
numerically introducing a spatial lattice with lattice constant $a$ which 
accordingly has to be chosen as $a\sim \varepsilon^{-1/4}$. 
We show that everything works out exactly in this manner.


This paper is organized as follows.
In Sec.~\ref{Sec:CYM}, we explain the equations of motion
in the CYM theory, and how we can obtain the eigenvalues
of the Hessian in CYM.
In Sec.~\ref{Sec:Resuls}, we show the eigenvalue distribution of the Hessian
and its time evolution.
Next we show the long-time evolution in terms of the maximum Lyapunov exponent.

\section{Theoretical background}
\label{Sec:CYM}

\subsection{Chaotic dynamics of Yang-Mills fields}

In this section, following a brief review of previous results, 
we discuss the method we use to analyze the complexity evolution
in the classical Yang-Mills theory for an intermediate time duration.
We first introduce the intermediate Lyapunov exponent which is applicable
to general cases, and apply it to the classical Yang-Mills evolution.

The chaotic properties of the classical evolution of Yang-Mills fields has been 
known and studied for a long time \cite{Biro:1994bi}. Chaos was first observed 
in the infrared limit of the Yang-Mills theory \cite{Matinyan:1986nw}; later it 
was shown to exist also in the compact lattice version of the classical Yang-Mills
theory \cite{Muller:1992iw}. The maximal global Lyapunov exponent may be
related to the plasmon damping rate of the thermal pure Yang-Mills plasma 
\cite{Biro:1994sh}.

The global KS entropy of the compact lattice gauge theory (i.e. the rate of entropy growth close
to thermal equilibrium) was shown to
be extensive, i.e.~proportional to the lattice volume \cite{Gong:1993xu},
and the ergodic properties of the compact SU(2) lattice gauge theory
were investigated numerically in detail by Bolte {\em et al.}~\cite{Bolte:1999th}.

Since we are here not interested in the ergodic properties of the classical 
nonabelian gauge theory, but in its dynamical properties far off equilibrium,
we will mostly make use of the non-compact formulation of the lattice gauge
theory. In the following, we set the stage for our investigation by discussing
three different kinds of instability exponents, which capture different aspects
of the dynamics of a nonlinear system with many degrees of freedom, such
as the classical Yang-Mills field.

\subsection{Local and intermediate Lyapunov exponents}


For a simple ``roll-over'' transition, $H=p^2/2-\lambda^2x^2/2$, 
we have one positive and one negative Lyapunov exponents,
$\lambda$ and $-\lambda$,
which characterize both, the kinetic instability and the entropy production.
This is understood in the matrix form as follows.
For a classical trajectory, $X(t)=(x(t), p(t))^T$, we consider a second trajectory
which differs a little in the initial condition.
The equations of motion for the tangent vector $\delta X(t) = (\delta x(t), \delta p(t))^T$
are written as,
\begin{align}
\dot{X}(t)
&=\begin{pmatrix}
0 & 1 \\
-1 & 0 \\
\end{pmatrix}
\begin{pmatrix}
H_x\\
H_p\\
\end{pmatrix}
\ ,\\
\delta \dot{X}(t)=&\
\begin{pmatrix}
0 & 1 \\
-1 & 0 \\
\end{pmatrix}
\begin{pmatrix}
H_{xx} & H_{xp} \\
H_{px} & H_{pp} \\
\end{pmatrix}
\delta X(t)
\ ,
\end{align}
where we have introduced short-hand notations, $H_x=\partial H/\partial x$,
$H_{xp}=\partial^2 H/\partial x\partial p$, and so on.
For an inverted harmonic oscillator, we put
$H_{xx}=-\lambda^2$, $H_{pp}=1$, and find,
\begin{align}
\delta \dot{X}(t)
=A \begin{pmatrix}
\lambda & 0 \\
0 & -\lambda \\
\end{pmatrix}
A^{-1}\delta X(t)
\ ,\quad
A=
\begin{pmatrix}
1 & -1 \\
\lambda & \lambda \\
\end{pmatrix}
\ .
\end{align}
This leads to an exponential expansion in the direction of $\lambda x +p$
and an exponential contraction in the direction of $-\lambda x +p$.
The entropy production rate in this simple case was analyzed
by KMOS, who found to be given by $dS/dt \to \lambda$ for $t \to \infty$.

In the case of many degrees of freedom, a similar structure will appear as
\begin{align}
\delta \dot{X}(t) =
\begin{pmatrix}
H_{px} & H_{pp} \\
-H_{xx} & -H_{xp} \\
\end{pmatrix}
\delta X(t)
\equiv \Hess(t) \delta X(t) \ .
\label{Eq:EOM}
\end{align}
Now the second derivatives should be regarded as matrices,
e.g. $(H_{xx})_{ij}=\partial^2H/\partial x_i \partial x_j$.
We will refer to the matrix of second derivatives, ${\cal H}$, as the 
Hessian in this paper.
The eigenvalues $\LLE$ of $\Hess$ are referred to as the local Lyapunov exponents (LLE).
The LLE plays the role of a ``temporally local'' Lyapunov exponent,
which specifies the departure of two trajectories in a short time period.

If $\Hess$ is constant, i.e.~in the absence of mode coupling, the LLEs 
are identical with the Lyapunov exponents, and the KS entropy is defined as 
the sum of positive LLEs.
In general, however, for a system with many degrees of freedom,
stable and unstable modes couple with each other.
Thus, the LLE does not generally agree with the Lyapunov exponent in a long time period.
In order to discuss the exponentially growing behavior of the fluctuation,
we introduce the intermediate Lyapunov exponent (ILE).

We can formally solve the equation of motion (\ref{Eq:EOM}) 
for a finite time period $\Delta t$ as,
\begin{align}
\delta X(t+\Delta t)
=\,&U(t,t+\Delta t) \delta X(t)
\ ,\\
U(t,t+\Delta t)
=\,& {\cal T} \left[\exp\left( \int_t^{t+\Delta t} \Hess(t+t') dt' \right)	\right]
\ ,
\end{align}
where ${\cal T}$ denotes the time ordered product.
Numerically, we can obtain the time-evolution operator $U$ by the Trotter formula, 
\begin{align}
U(t,t+\Delta t)=\,& {\cal T} \prod_{k=1,N}
U(t_{k-1},t_k)
\nn
\simeq\,& {\cal T} \prod_{k=1,N}
\left[1+ \Hess(t_{k-1}) \delta t
\right]
\ ,
\label{Eq:Trotter2}
\end{align}
where $\delta t = \Delta t/N$.
We diagonalize the time evolution matrix $U$ and define the ILEs as,
\begin{align}
U_D(t,t+\Delta t) =
\mathrm{diag}
(e^{\lambda^\mathrm{ILE}_1 \Delta t}, e^{\lambda^\mathrm{ILE}_2 \Delta t}, \ldots).
\end{align}
Liouville's theorem dictates that the determinant of the time evolution matrix $U$ is unity,
and thus the sum of all positive and negative ILEs is zero.
After a long enough time for thermalization,
the distribution of the ILEs is expected to converge to that of the global Lyapunov
exponents (GLE):
\begin{align}
\lambda^\mathrm{ILE} \to
\begin{cases}
	\lambda^\mathrm{LLE}	& (\Delta t \to 0)\ ,\\
	\lambda^\mathrm{GLE}	&(\Delta t \to \infty)\ ,
\end{cases}
\end{align}

In general all three types of Lyapunov exponents, LLE, ILE, and GLE,
yield different results.
Here we are interested in the rapid growth of the coarse grained entropy
when the gauge field configuration is still far from equilibrium,
but has already had sufficient time to sample a significant fraction of phase space.
Our goal is not to calculate how the entropy grows when a configuration close
to equilibrium relaxes further; this can be calculated reliably in thermal quantum field theory. 
Instead, we focus below on the ILEs,
and estimate the KS entropy as
\begin{equation}
\frac{dS}{dt}=S_{KS}=\sum_{\lambda^\mathrm{ILE}_i>0} \lambda^\mathrm{ILE}_i
\ .
\end{equation}

\subsection{Classical Yang-Mills equation}


We consider the pure Yang-Mills theory in the temporal gauge,
which permits a Hamiltonian formulation. The continuum Hamiltonian
is given in terms of the physical chromoelectric and chromomagnetic fields, 
$E_i^a$ and $B_i^a=\varepsilon_{ijk}F_{jk}^a$, by
\begin{equation}
H = \frac{1}{2g^2}\int d^3x \left( \sum_{a,i}E_i^a(x)^2 + \frac{1}{2}\sum_{a,i,j}F_{ij}^a(x)^2 \right) .
\end{equation}
We now define the dimensionless variables on the lattice
with lattice spacing $a$ as (omitting vector and color indices)
\begin{align}
A^\mathrm{L} = aA ,
\quad
E^\mathrm{L} = a^2E ,
\quad
F^\mathrm{L} = a^2F .
\end{align}
The time variable is rescaled as
\begin{equation}
t^\mathrm{L} = t/a .
\end{equation}
The lattice spacing $a$ is thus scaled out, and the dimensionless
lattice Hamiltonian is defined as
\begin{equation}
H^\mathrm{L} = a g^2 H .
\end{equation}
Here we make use of the fact that a rescaling of the Hamiltonian (by $g^2$) 
does not affect the classical equations of motion. 
In the following we omit the superscript ``$\mathrm{L}$''.
The Hamiltonian on the lattice is 
\begin{align}
H=&\frac{1}{2}\sum_{x,a,i}E_i^a(x)^2 + \frac{1}{4}\sum_{x,a,i,j}F_{ij}^a(x)^2 
\ ,\\
F_{ij}^a(x) &= \partial_i A_j^a (x)-\partial_j A_i^a (x)
+ \sum_{b,c} f^{abc}A^b_i(x)A^c_j(x)
\ ,
\end{align}
where $\partial_i$ is the central difference operator in the $i$-direction, i.e., $\partial_i A(x) \equiv \{ A(x+\hat{i})-A (x-\hat{i}) \}/2$.

The classical equations of motion are given as,
\begin{align}
\dot{A}_i^a(x) &= E_i^a(x) \ ,\\
\dot{E}_i^a(x) &= \sum_{j} \partial_j F_{ji}^a(x)+ \sum_{b,c,j} f^{abc}A^b_j(x) F_{ji}^c(x)
\ .
\end{align}
There are two conserved quantities; total energy and charge.
In the numerical simulation, we check that the total energy is strictly conserved along the time evolution.
The charge conservation is expressed by non-Abelian Gauss' law,
\begin{align}
\sum_{i} \partial_i E_{i}^a(x)+ \sum_{b,c,i} f^{abc}A^b_i(x) E_{i}^c(x)=0
\end{align}
In the non-compact formalism, the lattice discritization violates the charge conservation in the magnitude of the field amplitudes.

The Hessian of CYM theory is written as
\begin{align}
\Hess
=
\begin{pmatrix}
H_{EA} & H_{EE} \\
-H_{AA} & -H_{AE} \\
\end{pmatrix}
\ ,
\end{align}
where the matrix elements are
\begin{align}
H_{EE}&=\delta^{ab}\delta_{ij}\delta_{x,y}
\ ,\\
H_{EA}&=H_{AE}=0
\ ,\\
H_{AA}&=\frac{1}{4}\delta^{ab}P+\frac{1}{2}\sum_c f^{abc} Q^{c}
+\sum_{cde} f^{acd}f^{bce} R^{de}
,
\end{align}
with
\begin{align}
P=&
-(\delta_{x+\hat{i},y+\hat{j}} -\delta_{x+\hat{i},y-\hat{j}}
-\delta_{x-\hat{i},y+\hat{j}} +\delta_{x-\hat{i},y-\hat{j}})
\nn
&+\delta_{ij}\sum_k (2\delta_{x,y}-\delta_{x+\hat{k},y-\hat{k}}
-\delta_{x-\hat{k},y+\hat{k}} )\\
Q^{c} =& 
A^c_i (y) (\delta_{x,y+\hat{j}}-\delta_{x,y-\hat{j}})
-A^c_j (x) (\delta_{x+\hat{i},y}-\delta_{x-\hat{i},y})\nonumber\\
&+\delta_{ij}\sum_k \{ A^c_k (x)+A^c_k (y) \} (\delta_{x+\hat{k},y}
-\delta_{x-\hat{k},y})
\nn
&+2F^c_{ij}(x)\delta_{x,y}\\
R^{de} =& \{ -A^e_i(x)A^d_j(x) + \delta_{ij}\sum_{k}A^d_k(x)A^e_k(x) \} \delta_{x,y}
\ .
\end{align}
On the $L^3$ lattice, the number of the eigenvalues is $6(N_c^2-1)L^3 $.

\subsection{Physical scale}

In order to fix the scale of the theory, we consider a physical volume $V=a^3L^3$
in which the gauge field is thermalized at temperature $T$.
The total energy is given by:
\begin{equation}
\label{eq:epsL}
\VEV{H} =V\Edns(T)
=\frac{\VEV{H^\mathrm{L}}}{g^2a}
=\frac{L^3}{g^2a}\,\Edns^\mathrm{L}
\ ,
\end{equation}
where $\Edns^\mathrm{L}=\VEV{H^\mathrm{L}}/L^3$ is the energy per site, i.e., the energy density in lattice units.


A classical Yang-Mills theory on the lattice
is a classical system of $2L^3(N_c^2-1)$ oscillators
and has the thermal energy density
\begin{eqnarray}
\Edns^\mathrm{L} &=& 2(N_c^2-1) \frac{1}{L^3} \sum_{\bf k} |{\bf k}|
\frac{T^\mathrm{L}}{|{\bf k}|}
\nn 
& =& 2(N_c^2-1) C_\mathrm{L} T^\mathrm{L}
\ ,
\end{eqnarray}
where $C_\mathrm{L} = \sum_{\bf k}/L^3$ is a numerical coefficient
of order unity. 
The physical energy density of the lattice theory is
\begin{equation}
\label{eq:eps-cl}
\Edns_\mathrm{cl}(T) = \frac{\Edns^\mathrm{L}}{a^4g^2} = 2(N_c^2-1) C_\mathrm{L} \frac{T}{a^3}
\ ,
\end{equation}
where we have used the relation $T^\mathrm{L}=ag^2T$.
On the other hand, the energy density in the weakly interacting thermal 
quantum Yang-Mills theory is
\begin{eqnarray}
\Edns(T) &=& 2(N_c^2-1) \int \frac{d^3k}{(2\pi)^3} \frac{|{\bf k}|}{e^{|{\bf k}|/T}-1}
\nn 
& =& 2(N_c^2-1) \frac{\pi^2}{30} T^4 
\ .
\end{eqnarray}
The classical theory only applies to those modes of the continuum theory which
are highly occupied and for which the quantum corrections are not too large.
This condition imposes a lower limit on the lattice spacing of the classical theory.
One can either argue that the two expressions for the energy density should 
coincide, or that $a$ is the screening length of the corresponding quantum 
field theory.
In both cases this leads to the relation
\begin{equation}
\label{eq:Ta}
a \geq \frac{\theta}{T} 
\ ,
\end{equation}
where $\theta$ is a numerical constant of order unity
and $T \sim \varepsilon^{1/4}$ is introduced as a measure of 
the energy density.
(For example, $\varepsilon_\mathrm{cl}=\varepsilon$
leads to $a=(30C_L/\pi^2)^{1/3}/T \simeq 1.45/T$.)

The KS entropy growth 
rate, i.e.~the sum of all positive Lyapunov exponents, in lattice units is given by
\begin{equation}
S_\mathrm{KS}^\mathrm{(L)} = c_\mathrm{KS} L^3 (\Edns^\mathrm{L})^{1/4}
\ .
\end{equation}
The KS entropy density in lattice 
units is thus
\begin{equation}
s_\mathrm{KS}^\mathrm{(L)} = c_\mathrm{KS} (\Edns^\mathrm{L})^{1/4}
\ .
\end{equation}
The equilibrium entropy density of the classical Yang-Mills theory in the
continuum with ultraviolet cut-off, is according to (\ref{eq:eps-cl}),
\begin{equation}
s_\mathrm{eq}(T) = \frac{4}{3} \frac{\Edns_\mathrm{cl}}{T} 
= 2 (N_c^2-1) \frac{4\,C_\mathrm{L}}{3\,a^3}
\ ;
\end{equation}
the same result in lattice units is
\begin{equation}
s_\mathrm{eq}^\mathrm{(L)} = a^3 s_\mathrm{eq}(T) 
= 2 (N_c^2-1) \frac{4\,C_\mathrm{L}}{3}
\ ;
\end{equation}
The equilibration time in lattice units is thus
\begin{equation}
\tau_\mathrm{eq}^\mathrm{(L)} = 
\frac{s_\mathrm{eq}^\mathrm{(L)}}{s_\mathrm{KS}^\mathrm{(L)}} 
= 2 (N_c^2-1) \frac{4\,C_\mathrm{L}}{3\,c_\mathrm{KS}} (\Edns^\mathrm{L})^{-1/4}
\ .
\end{equation}
Finally, the physical equilibration time is
\begin{eqnarray}
\tau_\mathrm{eq} &=& \tau_\mathrm{eq}^\mathrm{(L)} a 
\geq \frac{\tau_\mathrm{eq}^\mathrm{(L)}\theta}{T}
\nn
&=&  2 (N_c^2-1) \frac{4\,C_\mathrm{L}\theta}{3\,c_\mathrm{KS}T} (\Edns^\mathrm{L})^{-1/4} 
\ .
\label{Eq:taueq}
\end{eqnarray}


\section{Classical Yang-Mills evolution}
\label{Sec:Resuls}

\subsection{Lyapunov exponents}
\label{Sec:Resuls-A}


We first discuss the Lyapunov exponents obtained by the numerical simulations of SU(2) CYM systems.
Initial conditions are prepared with $E^a_i(x)=0$ and random $A^a_i(x) \neq 0$.
To see the chaotic time evolution, we measured the ``distances'' between two gauge configurations:
\begin{align}
\label{eq:DEE}
D_{EE} &= \sqrt{ \sum_{x} \{ \sum_{a,i} E_{i}^a(x)^2 - \sum_{a,i} {E'}_{i}^{a}(x)^2 \}^2 } \ , \\
\label{eq:DFF}
D_{FF} &= \sqrt{ \sum_{x} \{ \sum_{a,i,j} F_{ij}^a(x)^2 - \sum_{a,i,j} {F'}_{ij}^{a}(x)^2 \}^2 } \ .
\end{align}
The two gauge configurations are set to be very close to each other at the initial time $t=0$.

\begin{figure}[tb]
\PSfig{8cm}{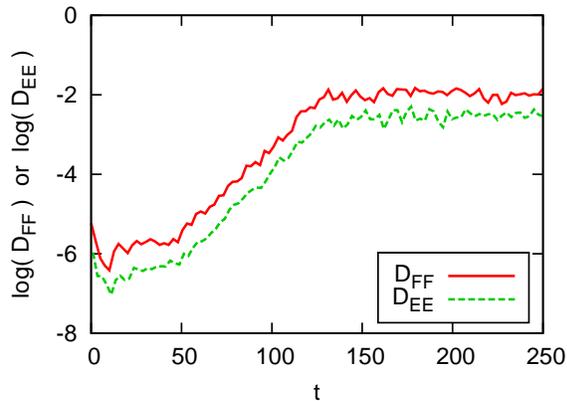}
\caption{
Time evolution of the distance in SU(2) simulation
on $4^3$ lattice.
All scales are given in the lattice unit.
}\label{Fig:AY1}
\end{figure}

\begin{figure}[tb]
\PSfig{8cm}{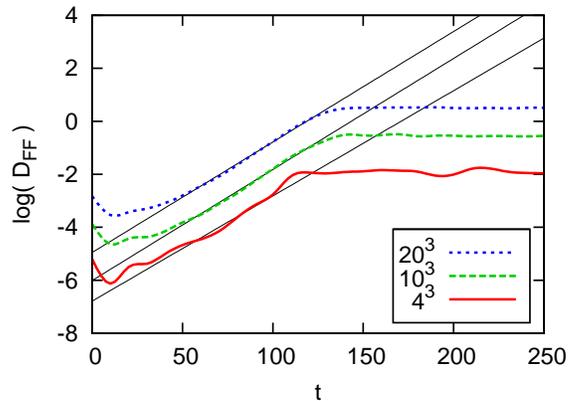}
\caption{
Time evolution in SU(2) simulation
on $4^3, 10^3$, and $20^3$ lattices with the same energy density.
}\label{Fig:AY2}
\end{figure}

In Fig.~\ref{Fig:AY1}, we show the numerical results on a $4^3$ lattice.
The energy density is $\varepsilon = 0.014$.
After a short time, the distance of two trajectories start to deviate,
and exponentially grows in the intermediate time region ($50 < t < 120$).
Later it saturates to a maximum value ($t > 120$).
The exponential growth rate of the distance, i.e., the linear slope of $\ln D_{FF}$, in the intermediate time region is $\lambda_D \sim 0.04$.
This growth rate $\lambda_D$ is governed by the maximum Lyapunov exponent for a finite time period.
In Fig.~\ref{Fig:AY2}, we show the lattice size dependence of the time evolution.
Apart from the irrelevant constant, the time evolution is almost insensitive to the lattice size.
This is consistent with the expectation that
the present lattice calculation simulates a piece of hot matter
occupying a much larger volume.

\begin{figure*}[tb]
\PSfig{8cm}{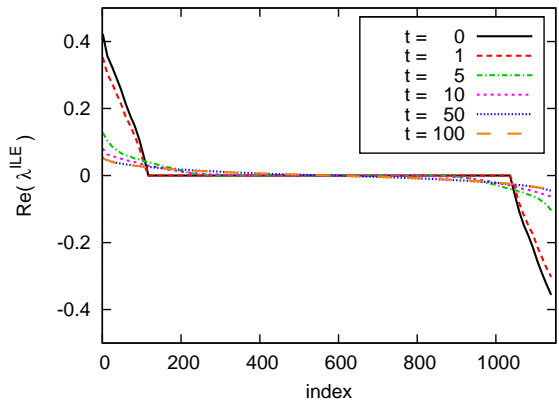}
\PSfig{8cm}{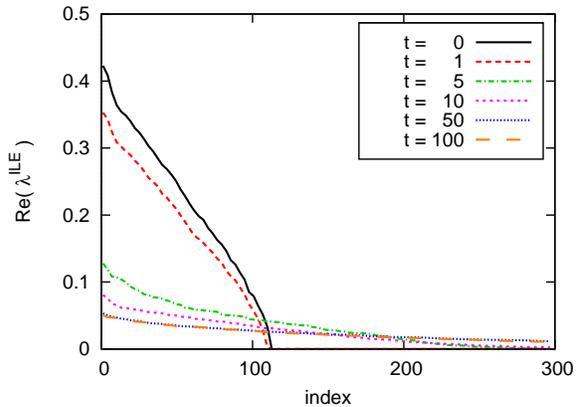}
\caption{
Distribution of the intermediate Lyapunov exponents.
The total number of the eigenvalues is 1152.
The right panel is a closeup of the largest 300 eigenvalues.
}\label{Fig:AY3}
\end{figure*}

We calculated the ILEs by using the Trotter formula (\ref{Eq:Trotter2}).
In the practical calculation, we have adopted the following
expression,
\begin{align}
1 + {\cal H}\delta t
\simeq  
\begin{pmatrix}
1 & \delta t\\
-H_{AA}\delta t & 1 - H_{AA}(\delta t)^2
\end{pmatrix}
,
\end{align}
which contains an ${\cal O}(\delta t^2)$ term and coincides with
$1 + {\cal H}\delta t$ up to ${\cal O}(\delta t)$.
The determinant of this matrix is equal to unity and thus protects the 
symplectic property of the evolution.
The eigenvalues are real or pure imaginary.
These eigenmodes correspond to the exponentially growing or damping mode and the oscillating mode, respectively.
Since Liouville's theorem ensures the sum of the ILEs is zero, the positive and negative ILEs should appear in a pairwise manner.

We show the ILE distribution in Fig.~\ref{Fig:AY3}.
The gauge configuration is the same as in Fig.~\ref{Fig:AY1}.
The distribution at $t=0$ corresponds to the LLEs of the initial condition.
A positive (negative) LLE corresponds to the temporally local negative (positive) potential curvature, and the maximum LLE is larger than $\lambda_D$.
Within a short time period ($0<t<5$), the maximal ILE rapidly decreases and the number of positive ILEs increases. 
As the distribution of ILEs no longer evolves for $t>50$, the KS entropy is, therefore, also constant for $t>50$.
In this time region, the maximum ILE is $\lambda_\mathrm{max}^\mathrm{LLE} \sim 0.04$, which is close to $\lambda_D$.
This fact means that the ILE does correspond to the growth rate for a finite time period.

\subsection{Growth of low- and high-momentum amplitudes}


Since the classical lattice theory is not ultra-violet (UV) safe,
the energy is exhausted in this limit mostly by UV modes,
which are sensitive to the lattice cutoff.
We note that the classical theory at nonzero temperature has no 
well-defined continuum limit; e.g., the Rayleigh-Jeans formula 
gives an energy density that diverges in that limit.
To wit, the thermal classical Yang-Mills
theory on a lattice has no 
well defined continuum limit and the choice of lattice constant
has physical significance, as discussed above.



In order to examine further the role of UV modes
we discuss next the momentum spectra of the gauge field $\tilde{A}(p)$.
We performed SU(2) simulations on a $N_s^3=16^3$ lattice
with the energy density $\varepsilon = 0.014$,
which is the same setup as that in Sec.~\ref{Sec:Resuls-A}.
The distance $D_{FF}$ then exhibits a similar behavior as Fig.~\ref{Fig:AY1}.

\begin{figure}[bt]
\PSfig{0.9\linewidth}{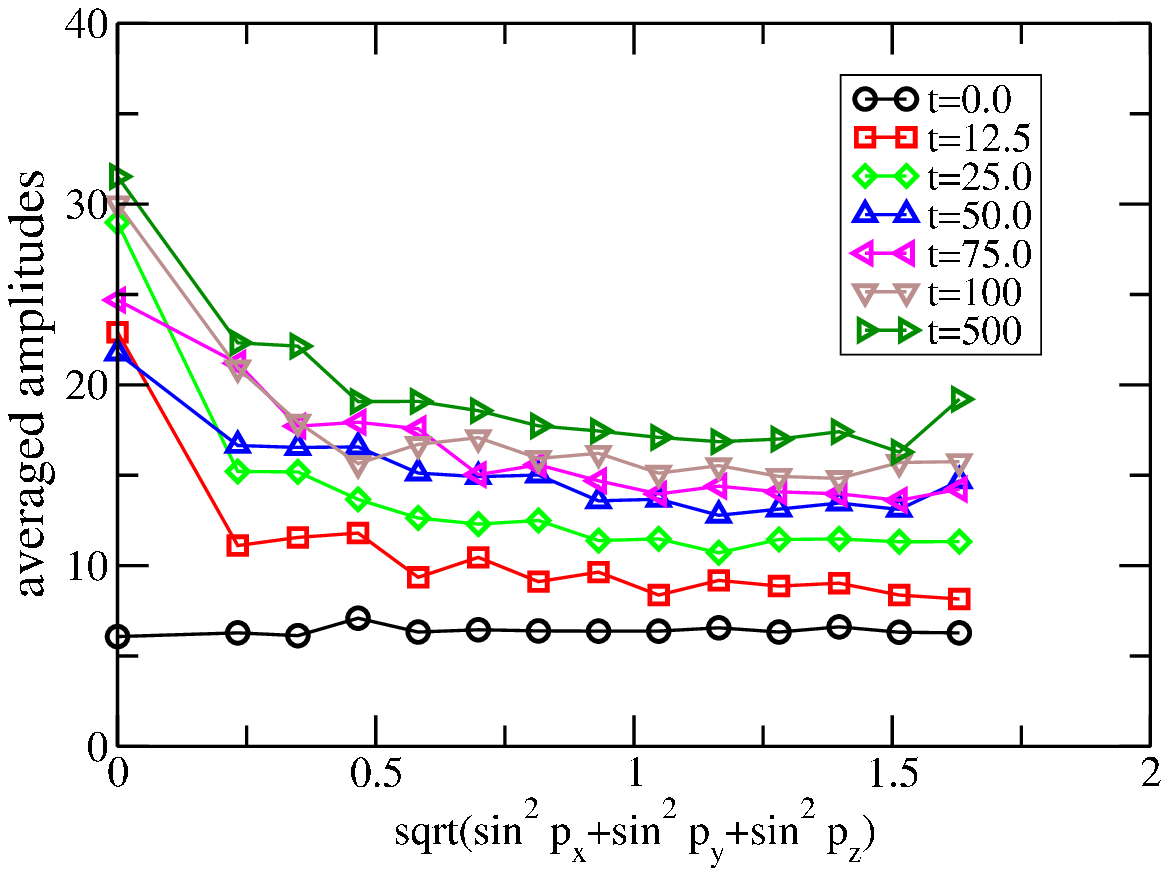}%
\caption{
The spectrum of gauge fields $\tilde{A}(|p|)$
for different times.
}\label{Fig-100308-2}
\end{figure}

The gauge-field's spectra $\tilde{A}(p)$
are obtained with 3-dim Fourier transformation of $A(x)$,
where momenta 
$p_\mu (\mu = 1, 2, 3)$ range from $-(N_s/2-1)$ to $N_s/2$
times $2\pi/(N_sa)$.
(Note that due to our definition of the Laplacian, which extends to 
$x\pm 2a$ we get a factor $2\pi/N_s$ rather than $\pi/N_s$). 

We average $|\tilde{A}_\mu(p)|$ over direction and color.
and show the time evolution of the spectrum $\tilde{A}(p)$ of $A(x)$
in Fig.~\ref{Fig-100308-2},
where the spectra $|\tilde{A}(p)|$ are plotted as functions of 
$\sqrt{\sin^2p_1+\sin^2p_2+ \sin^2p_3}$.
Due to discretization of space one encounters doublers, which is why 
only half of the Brillouin zone is plotted in Fig.~\ref{Fig-100308-2}.

At $t = 0$, $A(x)$ is randomly distributed and hence the $\tilde{A}(p)$
are almost independent of $|p|$.
After a short time the IR modes are strongly excited.
and they dominate the exponential growth of the distance between gauge field 
configurations, as was expected.




At low momenta our results approach the
classical equilibrium (equipartition) distribution
explains the tendency of our results at low momenta,
but it is not completely reached even in the IR modes 
at the last stage of the exponential growth ($t \sim 150$).


Our results show, in addition, that at earlier times 
($t<50$) one has IR modes with very rapid growth,
which appear to be the modes associated with the largest Lyapunov
exponents. This would fit the usual assumption of a bottom-up 
thermalization~\cite{Baier:2000sb}, except that it is rather a pre-thermalization,
because phase space is filled rapidly, but the full approach towards equilibrium 
sets in only with the linear phase, i.e.~for $t> 50$.

\subsection{Equilibration time of SU(3) Yang-Mills theory}
\label{Subec:SU3}

\begin{figure}[tb]
\PSfig{8cm}{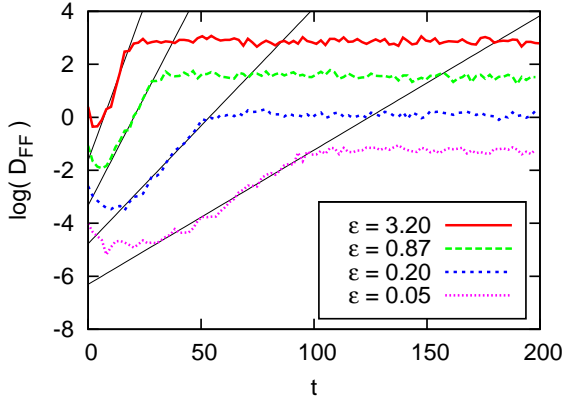}
\caption{Time evolution in SU(3) simulation on a $4^3$ lattice.
}\label{Fig:AY4}
\end{figure}

\begin{figure}[tb]
\PSfig{8cm}{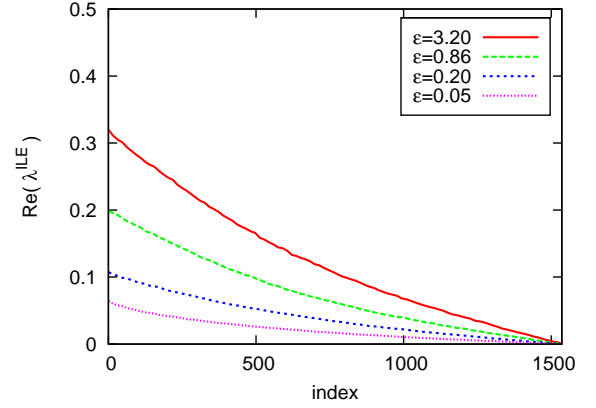}
\caption{Distribution of the intermediate Lyapunov exponents in SU(3) simulations.
}\label{Fig:AY5}
\end{figure}

Next, we discuss the SU(3) CYM theory.
In Fig.~\ref{Fig:AY4}, we show the time evolution of $D_{FF}$ in SU(3) simulation on a $4^3$ lattice for several energy densities.
By changing the initial amplitude of $A_i^a(x)$, we calculated time evolutions with different energy densities.
In Fig.~\ref{Fig:AY5}, we show the ILE distributions after a long time period, which no longer change along time evolution.
Only the positive eigenvalues are shown.
These are qualitatively the same as the SU(2) simulations.

\begin{figure}[tb]
\PSfig{7cm}{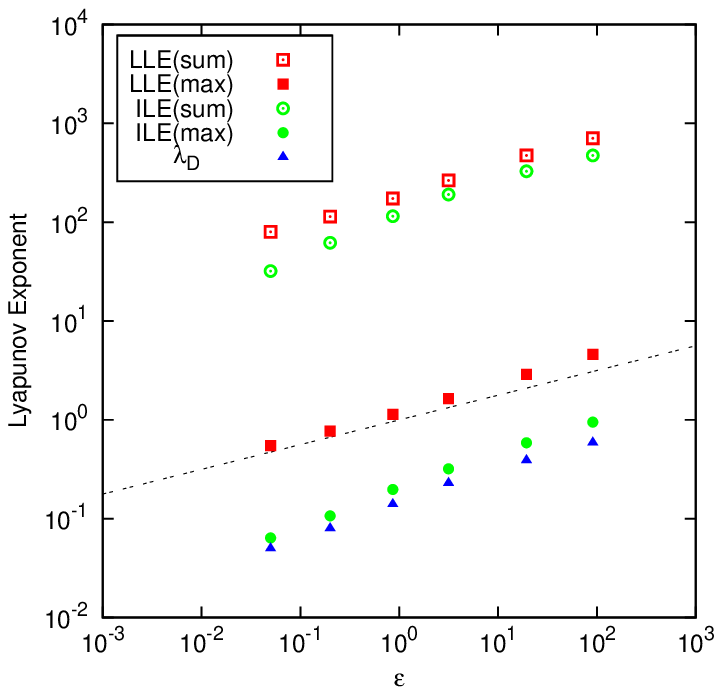}
\caption{The SU(3) results of the Lyapunov exponents, $\lambda_D$,
$\lambda_\mathrm{max}^\mathrm{LLE}$,
$\lambda_\mathrm{sum}^\mathrm{LLE}$,
$\lambda_\mathrm{max}^\mathrm{ILE}$,
and $\lambda_\mathrm{sum}^\mathrm{ILE}$.
The broken line is $\Edns^{1/4}$.
}\label{Fig:AY6}
\end{figure}

In Table~\ref{Table:SU3} and Fig.~\ref{Fig:AY6}, we show the SU(3) results of the Lyapunov exponents:
the exponential growth of the distance $\lambda_D$,
the largest LLE $\lambda_\mathrm{max}^\mathrm{LLE}$,
the sum of the positive LLEs $\lambda_\mathrm{sum}^\mathrm{LLE}$,
the largest ILE $\lambda_\mathrm{max}^\mathrm{ILE}$,
and the sum of the positive ILEs $\lambda_\mathrm{sum}^\mathrm{ILE}$.
As discussed in the previous section, the Lyapunov exponents should scale as $\Edns^{1/4}$ because of the conformal invariance.
As shown in Fig.~\ref{Fig:AY6}, $\lambda_\mathrm{max}^\mathrm{LLE}$ and $\lambda_\mathrm{sum}^\mathrm{LLE}$ are indeed proportional to $\Edns^{1/4}$.
Other Lyapunov exponents slightly deviate from this scaling.
This is because the change of the field amplitude is not exactly the conformal transformation, e.g., the dimensionless ratio of the electric energy density to the magnetic energy density is changed.
Numerically, however, the best-fit prefactor is not much affected by the difference of the exponent in the following accuracy.

After all, we extract the Lyapunov exponent as a function of the energy density from this approximate scaling.
By fitting, we find that the numerical prefactors are
\begin{align}
\lambda_D &\simeq 0.1\times \Edns^{1/4} \ ,\\
\lambda_\mathrm{max}^\mathrm{LLE} &\simeq 1 \times \Edns^{1/4} \ ,\\
\lambda_\mathrm{sum}^\mathrm{LLE}/L^3 &\simeq 3 \times \Edns^{1/4} \ ,\\
\lambda_\mathrm{max}^\mathrm{ILE} &\simeq 0.2\times \Edns^{1/4} \ ,\\
\lambda_\mathrm{sum}^\mathrm{ILE}/L^3 &\simeq 2\times \Edns^{1/4} \ .
\end{align}
Thus, the KS entropy density is
\begin{equation}
s_\mathrm{KS} = \lambda_\mathrm{sum}^\mathrm{ILE}/L^3 \simeq 2\times \Edns^{1/4} 
= c_{KS} \times \Edns^{1/4}  \ .
\end{equation}

\begin{table}[bt]
\renewcommand{\tabcolsep}{0.5pc} 
\renewcommand{\arraystretch}{1} 
\caption{The Lyapunov exponents in the SU(3) CYM theory.
}\label{Table:SU3}
\begin{tabular}{ccccccc}
\hline
\hline
$L^3$
& $\varepsilon$ 
& $\lambda_D$
& $\lambda^\mathrm{LLE}_\mathrm{max}$
& $\lambda^\mathrm{LLE}_\mathrm{sum}$
& $\lambda^\mathrm{ILE}_\mathrm{max}$
& $\lambda^\mathrm{ILE}_\mathrm{sum}$\\
\hline
$4^3$& 0.05 & 0.05 & 0.55 & 80  & 0.06 & 32  \\ 
$4^3$& 0.20 & 0.08 & 0.77 & 114 & 0.11 & 62  \\
$4^3$& 0.86 & 0.14 & 1.14 & 174 & 0.20 & 115 \\
$4^3$& 3.16 & 0.23 & 1.64 & 265 & 0.32 & 191 \\
$4^3$& 19.4 & 0.39 & 2.89 & 474 & 0.59 & 328 \\
$4^3$& 90.9 & 0.59 & 4.60 & 708 & 0.95 & 474 \\
\hline
\hline
\end{tabular}
\end{table}

From this result, we can evaluate the equilibration time of the SU(3) CYM theory.
We take
$\theta \simeq (30C_L/\pi^2)^{1/3} \simeq 1.45$ and
$C_{\rm L}\simeq 1$ 
as a typical case.
Then we get $T^\mathrm{L} = ag^2T=g^2(30/\pi^2)^{1/3} \simeq 6$
and $\varepsilon^{\rm L} = 2(N_c^2-1)C_\mathrm{L}T^\mathrm{L} \simeq 90$.
Inserting these numbers into Eq.~(\ref{Eq:taueq}), we obtain
\begin{equation}
\tau_\mathrm{eq} \simeq \frac{5}{T}~+~ \tau_{\rm delay}
\ .
\end{equation}
Here $\tau_{delay}$ was introduced to take the initial phase into account, 
in which 
$D_{FF}$ is more or less constant, because the strongly growing modes are 
not yet relevant.  
One would expect that $\tau_{\rm delay}$ fulfills approximately
\cite{Biro:1993qc}
 
\begin{equation}
\frac{1}{6(N_c^2-1)L^3} e^{\lambda_{\rm max}\tau_{\rm delay}} ~\approx ~ 1
\end{equation}
which is indeed in good agreement with our numerical findings.
When $T \simeq 350$ MeV, the equilibration time is
$\tau_\mathrm{eq} \simeq 3$ fm/$c$, with a systematic uncertainty which 
can easily account for a factor of two.\\
If entropy is produced by very rapid processes, especially 
by decoherence, before the non-linear dynamics 
analysed by us reaches the phase of linear entropy growth, the 
real thermalization time is correspondingly shorter. In 
\cite{Fries:2008vp} this decoherence entropy was estimated to be roughly 
1/3 of what is needed by the hydrodynamical initial conditions. 
This is consistent with results obtained in \cite{albacete} in 
$k_{\perp}$ factorized perturbation theory. In that calculation
the full observed particle number at
central rapidities 
is only reached 
from decohering the Color Glass Condensate
after introducing a normalization factor.  
Without that factor one would obtain between one half and one quarter 
of the total particle number. 
On the other hand, simulations of the combined 
decoherence and non-linear dynamics stage of the glasma in a longitudinally 
expanding, boost-invariant geometry reaches 
80\% of the final particle number \cite{lappi}. All of this indicates that 
it is probably a good guess to assume that non-linear gauge field dynamics 
has to generate about 2/3 of the entropy required by thermal equilibrium
and that the thermalization time is thus rather of the order of 2 fm/c.

\section{Summary}
\label{Sec:Summary}


The main aim of this paper is to understand
the fast thermalization deduced for high energy heavy ion collissions, 
which is, in fact, one of the largest unexplained puzzles in RHIC physics.
We argue that entropy generation plays the key role in this context.

The overall picture of entropy generation in heavy ion collins is involved.
While some part of the entropy is produced from the decoherence
at very early times, i.e. times of order  $1/Q_s$, see~\cite{Fries:2008vp},
most of entropy required by the initial condition for the
hydrodynamic phase must be generated within the first fm/c 
by nonequilibrium gluon dynamics. 
Entropy generation of quantum systems always requires coarse graining.
Coarse-graining in turn relates it to the Lyapunov exponents, see
~\cite{Fries:2008vp}. As the latter can be studied 
in the corresponding classical gauge theories so can entropy production 
in total.

More precisely, 
the entropy production rate in classical systems is given by
the Kolmogorov-Sina\"i (KS) entropy,
defined as the sum of positive Lyapunov exponents.
(The KS entropy describes the entropy production also in quantum systems
when the coarse graining is introduced
with a minimum wave packet~\cite{Kunihiro:2008gv}.)

We have investigated classical Yang-Mills (CYM) dynamics
in the noncompact $(A,E)$ scheme. We started from random initial conditions 
and studied the resulting spectrum of Lyapunov exponents.
We found that their properties change with time in a characteristic manner 
and identified three distinct regimes: A short time regime, in which 
the system has not yet sampled a large fraction of phase space, a late time 
regime 
in which the system is already close to thermal equilibrium and has sampled 
basically all of phase space, and an intermediate regime which is dominated by 
non-linear gauge field dynamics.

We have developed a method, making use of Trotter formula, 
to evaluate the Lyapunov exponent
in the intermediate time scale (intermediate Lyapunov exponent; ILE),
which is the relevant time scale for the problem of thermalization 
in heavy ion collisions, 
and determined the entropy production rate (Kolmogorov-Sina\"i entropy).
The obtained equilibration time scales as
$\tau_\mathrm{eq} \propto \varepsilon^{-1/4} \propto 1/T+\tau_{\rm delay}$,
where $\varepsilon$ is the energy density
and $\tau_{\rm delay}$ is the typical time to reach the
intermediate time range, which we also determined.
In total the thermalization time is 
around 2 fm$/c$ for $T = 350~\mathrm{MeV}$, if one assumes
that 1/3 of the required entropy is generated by decoherence,
with rather substantial systematic uncertainties. 
The most important source of uncertainty is related to the choice
of lattice constant $a$.
Since CYM has conformal symmetry, the physical scale setting is provided 
by the discretization scale $a$ which thus acquires physical importance. 
The choice of $a$ is not free of ambiguities. 
Different arguments all lead to the form $a=c\varepsilon^{1/4}$ with 
a constant $c$ of the order of one but varying within a factor of two.

One also finds that the $\varepsilon$ dependence of $a$ is crucial
to obtain the correct power scaling for all quantities of interest from CYM.

In the course of these investigations it was crucial to understand the 
qualitative differences between the different time ranges and corresponding 
Lyapunov spectra, which also allows to reconcile previously not understood
observations \cite{Mike_Yasushi}.

A thermalization time of roughly 2 fm$/c$ is somewhat 
larger than the phenomenologically preferred value. 
However, the inclusion of quarks will reduce this number 
and could well bring it into the phenomenologically preferred range.
In addition 
strong electric field in the initial
condition together with the magnetic field, and longitudinal (Bjorken)
expansion may promote faster equilibration.

\section*{ACKNOWLEDGMENTS}
We acknowledge helpful discussions with A.~Dumitru, Y.~Nara, and
M.~Strickland.
Part of this work was done during the Nishinomiya-Yukawa Memorial Workshop
``High Energy Strong Interactions 2010''.
This work was supported by the Bundesministerium f\"ur Bildung und Forschung
in Germany, by the Global COE Program ``The Next Generation of Physics, Spun
from Universality and Emergence'' in Kyoto University, by the Yukawa
International Program for Quark-hadron Sciences in YITP, by the
Grant-in-Aid for Scientific Research in Japan [Nos. 20540265, 21740181, 
22105508 and 20$\cdot$363], and by the U.S. Department of Energy
[grant DE-FG02-05ER41367].
B.M.~and A.S.~thank YITP for its hospitality and support.
The numerical calculations were carried out on Altix3700 BX2 and SX8 at
YITP.

\end{document}